\documentclass[11pt]{article}
\usepackage{graphicx}

\newcommand{\BABARPubYear}    {08}
\newcommand{\BABARConfNumber} {009}
\newcommand{\SLACPubNumber} {13346}

\input pubboard/babarsym
 
\long\def\inst#1{\par\nobreak\kern 4pt\nobreak
    {\it #1}\par\vskip 10pt plus 3pt minus 3pt}

\RequirePackage{xspace}

\newcommand{\calH}{\ensuremath{{\cal H}}}
\newcommand{\pvec}{{\bf p}}

\newcommand{\calB}{\ensuremath{{\cal B}}}

\newcommand{\bfemsix}{\ensuremath{\calB(10^{-6})}}

\newcommand{\DE}{\ensuremath{\Delta E}}

\newcommand{\xf}{\ensuremath{{\cal F}}}

\newcommand{\thetaT}{\ensuremath{\theta_{\rm T}}}
\newcommand{\costhr}{\ensuremath{\cos\thetaT}}

\newcommand\etal{{\it et al.}}
\newcommand{\half}{\ensuremath{{1\over2}}}

\newcommand{\bfig}{\begin{figure}[htbpc!]}
\newcommand{\efig}{\end{figure}}
\newcommand\bef{\begin{figure}}
\newcommand\edf{\end{figure}}

\newcommand\sgline{\noalign{\vskip 0.10truecm\hrule\vskip 0.10truecm}}
\newcommand\beq{\begin{equation}}
\newcommand\eeq{\end{equation}}
\newcommand\bear{\begin{array}}
\newcommand\enar{\end{array}}
\newcommand\beqa{\begin{eqnarray}}
\newcommand\eeqa{\end{eqnarray}}
\newcommand\ben{\begin{enumerate}}
\newcommand\een{\end{enumerate}}

\newcommand{\UfourS}{\ensuremath{\Upsilon(4S)}}

\newcommand{\Kstz}{\ensuremath{\Kstarz}}

   \newcommand{\rhoz}{\ensuremath{\rho^0}}

   \newcommand{\aone}{\ensuremath{a_1}}
   \newcommand{\aonep}{\ensuremath{a_1^+}}

\newcommand{\BcacKstz}{\ensuremath{B^+ \to \aonep \Kstz}}
\newcommand{\rakst}{\ensuremath{0.7^{+0.5}_{-0.4} {}^{+0.7}_{-0.7}}}
\newcommand{\Rakst}{\ensuremath{(\rakst)\times 10^{-6}}}
\newcommand{\sakst}{\ensuremath{0.9}}
\newcommand{\ulakst}{\ensuremath{1.6}\xspace}
\newcommand{\Ulakst}{\ensuremath{\ulakst\times 10^{-6}}\xspace}
\newcommand{\ntotalpToy}{\ensuremath{15802}}
\newcommand{\resultFL}{\ensuremath{1.1 \pm 0.2}}
\newcommand{\resultY}{\ensuremath{55}}
\newcommand{\fitbias}{\ensuremath{27}}
\newcommand{\rakstprodBR}{\ensuremath{1.5^{+1.0}_{-0.9} {}^{+1.4}_{-1.4}}}
\newcommand{\RakstprodBR}{\ensuremath{(\rakstprodBR)\times 10^{-6}}}
\newcommand{\ulakstprodBR}{\ensuremath{3.3}\xspace}
\newcommand{\UlakstprodBR}{\ensuremath{\ulakstprodBR\times 10^{-6}}\xspace}

\begin{document}
{\pagestyle{empty}

\begin{flushright}
\babar-CONF-\BABARPubYear/\BABARConfNumber \\
SLAC-PUB-\SLACPubNumber \\
July 2008 \\
\end{flushright}

\par\vskip 5cm

\begin{center}
\Large \bf Search for $B^+$-meson decay to $a_1^+\Kstz$
\end{center}
\bigskip

\begin{center}
\large The \babar\ Collaboration\\
\mbox{ }\\
\today
\end{center}
\bigskip \bigskip

\begin{center}
\large \bf Abstract
\end{center}

We present the preliminary result of a search for the decay $B^{\pm} \ra a_1^{\pm} \Kstarz$.
The data, collected with the \babar\ detector at the Stanford Linear Accelerator
Center, represent 465 million \BB\ pairs produced in \epem\ annihilation at the
\FourS\ energy.  The result for the branching fraction is:
\begin{eqnarray*}
\calB(B^+ \to \aonep \Kstz) \times \calB(\aonep \to \pi^+ \pi^- \pi^+) =\Rakst \mathrm{,}
\end{eqnarray*}
corresponding to an upper limit at 90\%\ confidence level of \Ulakst.
The first error quoted is statistical, the second systematic.
\vfill
\begin{center}

Submitted to the 34$^{\rm th}$ International Conference on High-Energy Physics, ICHEP 08,\\
29 July---5 August 2008, Philadelphia, Pennsylvania.

\end{center}

\vspace{1.0cm}
\begin{center}
{\em Stanford Linear Accelerator Center, Stanford University, 
Stanford, CA 94309} \\ \vspace{0.1cm}\hrule\vspace{0.1cm}
Work supported in part by Department of Energy contract DE-AC02-76SF00515.
\end{center}

\newpage
}

\begin{center}
\small

The \babar\ Collaboration,
\bigskip

%
B.~Aubert,
M.~Bona,
Y.~Karyotakis,
J.~P.~Lees,
V.~Poireau,
E.~Prencipe,
X.~Prudent,
V.~Tisserand
\inst{Laboratoire de Physique des Particules, IN2P3/CNRS et Universit\'e de Savoie, F-74941 Annecy-Le-Vieux, France }
J.~Garra~Tico,
E.~Grauges
\inst{Universitat de Barcelona, Facultat de Fisica, Departament ECM, E-08028 Barcelona, Spain }
% ...

\end{center}\newpage

\section{INTRODUCTION}
\label{sec:Introduction}
Recent searches for decays of \B\ mesons to final states with an
axial-vector meson $a_1$ or $b_1$ and a pion or kaon have revealed
modes with
branching fractions that are rather large among charmless decays:
(15--35)$\times10^{-6}$ for $\B\ra a_1(\pi, K)$ 
\cite{BaBar_a1pi, BaBar_a1K}, and (7--11)$\times10^{-6}$ for a charged pion
and kaon in combination with a $b_1^0$ or a $\b_1^+$ meson
\cite{BaBar_b1h, conjugate}. On the other hand the experimental search
for $B^0 \to b_1^- \rho^+$ set an upper limit of $1.7\times10^{-6}$ 
at the 90\%\ confidence level for the branching fraction \cite{fordFPCP}, 
although a branching fraction of $25\times10^{-6}$ was expected \cite{ChengYang}.

The available theoretical estimates of the branching fractions of $B^+$
mesons to $\aonep \Kstz$ come from calculations based on na\"{i}ve
factorization \cite{Calderon}, and on QCD factorization
\cite{ChengYang}. The latter incorporates light-cone distribution
amplitudes evaluated from QCD sum rules, and predicts branching fractions
in quite good agreement with the measurements for $\B\ra a_1 \pi$ and
$\B\ra a_1 K$ \cite{BaBar_a1pi, BaBar_a1K}.
The expected branching fraction for \BcacKstz\ 
from na\"{i}ve factorization is $0.51 \times 10^{-6}$
and that from QCD factorization is
$9.7^{+4.9}_{-3.5}{}^{+32.9}_{-2.4} \times 10^{-6}$ with a prediction 
for the longitudinal polarization fraction $f_L = 0.38^{+0.51}_{-0.40}$.
The first theoretical error corresponds to the uncertainties due to the variation
of Gegenbauer moments, decay constants, quark masses, form factors and a 
$B$ meson wave function parameter. The second theoretical error corresponds to
the uncertainties due to the variation of penguin annihilation parameters~\cite{ChengYang}.
For the longitudinal polarization fraction, all errors are added in quadrature,
since the theoretical uncertainty is dominated uncertainties in the size of the
penguin-annihilation amplitude.
This mode is expected to be substantially enhanced
by penguin annihilation and thus it is important to study this mechanism.
In this paper we present the first search for the decay \BcacKstz.

\section{THE \babar\ DETECTOR AND DATASET}
\label{sec:babar}
The data for this measurement were collected with the \babar\
detector~\cite{BABARNIM} at the PEP-II asymmetric $e^+e^-$ collider
located at the Stanford Linear Accelerator Center.  An integrated
luminosity of 424 \invfb, corresponding to $(465\pm5)\times 10^6$ \BB\
pairs, was produced in \epem\ annihilation at the $\Upsilon (4S)$
resonance (center-of-mass energy $\sqrt{s}=10.58\ \gev$).
Charged particles from the \epem\ interactions are detected, and their
momenta measured by a combination of five layers of double-sided
silicon microstrip detectors and a 40-layer drift chamber both
operating in the 1.5~T magnetic field of the BaBar superconducting
solenoid. Photons and electrons are identified with a CsI(Tl)
electromagnetic calorimeter (EMC).  Further charged particle
identification (PID) is provided by the average energy loss ($dE/dx$) in
the tracking devices and by an internally reflecting ring imaging
Cherenkov detector (DIRC) covering the central region.

A detailed Monte Carlo program (MC) is used to simulate the \B\
meson production and decay sequences, and the detector response
\cite{geant}. 
Dedicated signal MC events for the decay
$B^+ \to \aonep \Kstarz$ with $\aone^+ \to \rho^0 \pi^+$
has been produced.
For the $\aone(1260)$ meson parameters we use a mass of $1230 \mevcc$
and a width of $400 \mevcc$. We account for the uncertainties of these resonance parameters 
in the determination of systematic uncertainties. The $\aonep \to \pi^-\pi^+\pi^+$ decay
proceeds mainly through the intermediate states $(\pi\pi)_\rho \pi$ and 
$(\pi\pi)_\sigma \pi$ \cite{PDG2006}. No attempt is made to separate contributions
of the dominant P wave $(\pi\pi)_\rho$ from the S wave $(\pi\pi)_\sigma$ in the channel 
$\pi \pi$. The difference in efficiency for the S wave and P wave cases is
accounted for as a systematic error.

\section{ANALYSIS METHOD}
\label{sec:Analysis}

We reconstruct \aonep\ candidates through the decay sequence
$\aonep\ra\rhoz\pip$ and $\rhoz\ra\pip\pim$. The other primary daughter
of the \B\ meson is reconstructed as $\Kstarz\ra\Kp\pim$. For the \rhoz, the
invariant mass of the pion pair is required to lie between 0.55 and 1.0 \gevcc.
For the \aone\ and $\Kstar$ we accept a range that includes sidebands. 
The \aone\ invariant mass is required to lie
between 0.9 and 1.8 \gevcc, while the \Kstar\ invariant mass is required to lie
between 0.8 and 1.0 \gevcc.
Secondary charged pion candidates from \aone\ and \Kstar\ decays are rejected if classified
as protons, kaons, or electrons by their DIRC, $dE/dx$, and EMC PID signatures. 
We reconstruct the \B\ meson candidate by combining the four-momenta of
a pair of primary daughter mesons, using a fit that constrains all
particles to a common vertex.
From the kinematics of \UfourS\ decay we determine the 
energy-substituted mass $\mes=\sqrt{\frac{1}{4}s-\pvec_B^2}$ and
energy difference $\DE = E_B-\half\sqrt{s}$, where $(E_B,\pvec_B)$ is
the \B\ meson four-momentum vector, and all values are expressed in the
\UfourS\ rest frame.
We require $5.25\ \gevcc <\mes<5.29\ \gevcc$ and $|\DE|<100\ \mev$.  
To reduce fake meson candidates we require a \B, \aone\ and \Kstar\
vertex $\chi^2$ probability $> 0.01$.

We also impose restrictions on the helicity-frame decay angle $\theta_{\Kstar}$
of the \Kstar\ mesons. The helicity frame of a meson is defined as
the rest frame of the meson with the $z$ axis along the direction of boost
to that frame from the parent rest frame.  For the decay
$\Kstar\ra K\pi$, $\theta_{\Kstar}$ is the polar angle of the
daughter kaon, and for $\aone\ra\rho\pi$, $\theta_{\aone}$ is the polar angle
of the normal to the \aone\ decay plane. We define $\calH_i = \cos(\theta_i)$, where
$i=(\Kstar,\aone)$.
Since many background candidates accumulate near $|\calH_{\Kstar}|=1$,
we require $-0.98 \le \calH_{\Kstar} \le 0.8$.

Backgrounds arise primarily from random combinations of particles in
continuum $\epem\ra\qqbar$ events ($q=u,d,s,c$).  We reduce these with
a requirement on the angle \thetaT\ between the thrust axis
\cite{thrust} of the \B candidate in the \UfourS\ frame and that of the
charged tracks 
and neutral calorimeter clusters in the rest of the event (ROE).
The distribution is sharply peaked near $|\costhr|=1$
for jet-like continuum events, and nearly uniform for \B\ meson decays.  The
requirement, which optimizes the expected signal yield relative to its
background-dominated statistical uncertainty, is $|\costhr|<0.8$.
\BB\ background arising from $b\ra c$ transitions is suppressed by 
removing events with $D$ meson candidates, reconstructed in the decays
$D^0 \to K^- \pi^+$ and $D^+ \to K^- \pi^+ \pi^+$, 
with an invariant mass within $\pm 0.02 \gevcc$ of the nominal mass value.

The number of events which pass the selection is \ntotalpToy.
Besides the signal events, these samples contain \qqbar\ 
(dominant) and \BB\ with $b\ra c$ combinatorial backgrounds, and a
fraction of other charmless \BB\ background modes.
The average number of candidates found per event in the selected
data sample is 1.5 (2.0 to 2.4 in signal MC depending on the polarization).
We choose the candidate that is most likely a signal decay, as judged from
the output of a neural network, where we use the  
$\rho$ meson mass and the vertex fit $\chi^2$ probabilities of \B, \aone\
and \Kstar\ candidates as input variables.

We discriminate further against \qqbar\ background with a
Fisher discriminant \xf\ \cite{fisher} that combines four variables: 
the polar angle of the $B$ candidate momentum and of the $B$ thrust axis
with respect to the beam axis in the \UfourS\ rest frame;
and the zeroth and second 
angular moments $L_{0,2}$ of the energy flow, excluding the $B$
candidate, with respect to the $B$ thrust axis.
The moments are defined by $ L_j = \sum_i
p_i\times\left|\cos\theta_i\right|^j,$ where $\theta_i$ is the angle
with respect to the $B$ thrust axis of a track or neutral cluster $i$,
$p_i$ is its momentum.

We obtain yields and the longitudinal polarization $f_L$ from an extended
maximum  likelihood (ML)
fit with the input observables \DE, \mes, \xf, the resonance masses
$m_{\aone}$ and $m_{\Kstar}$ and the helicity distributions $\calH_{\Kstar}$ 
and $\calH_{\aone}$.

Since the correlation between the observables in the selected data
and in MC signal events is small, we take the probability density function
(PDF) for each event to be a product of the PDFs for the individual
observables. Corrections for the effects of possible correlations are made
on the basis of MC studies described below.

We determine the PDFs for the signal and \BB\ background components from
fits to MC samples. We develop PDF parameterizations for the combinatorial
background with fits to the data from which the signal region
($5.26\ \gevcc <\mes<5.29\ \gevcc $ and $|\DE|<60\ \mev$) has been excluded. 

The \mes\ and \DE\ distributions are parametrized
as linear combinations of the so-called Crystal-Ball
function~\cite{CB} and Gaussian. In case of \mes\ for \qqbar and \BB\
background we use the threshold function
$x\sqrt{1-x^2}\exp{\left[-\xi(1-x^2)\right]}$, with argument
$x\equiv2\mes/\sqrt{s}$ and shape parameter $\xi$. This function is 
discussed in more detail in Ref.~\cite{PRD04}. In case of 
\DE\ for \qqbar and \BB\ background we use a polynomial function.
The PDFs for the Fisher discriminat ${\cal P}_j(\xf)$ are parametrized
as single or double Gaussian. The PDFs for the invariant masses of 
the \aone\ and \Kstar\ mesons are constructed
as linear combinations of relativistic Breit Wigner and polynomial functions.
We use a joint PDF ${\cal P}_j (\calH_{\Kstar}, \calH_{\aone})$  for the
helicity distributions, the signal component is parametrized as the product
of the ideal angular distribution in $\calH_{\Kstar}$ and
$\calH_{\aone}$ from Ref.~\cite{datta}, times an empirical acceptance 
function ${\cal G}(\calH_{\Kstar}, \calH_{\aone})$ while the helicity PDF
for the other components is the product of the helicity PDFs for 
$\calH_{\Kstar}$ and $\calH_{\aone}$. The $\calH_i$ distributions in case
of \qqbar and \BB\ background are based on Gaussian and polynomial functions.
We allow the most important parameters (first coefficient of the polynomial 
function for \DE, the invariant masses of the \aone\ and the \Kstar, and
the width of the Breit Wigner for the invariant masse of the \Kstar) for
the determination of the combinatorial background PDFs to vary in the fit, along with the
yields for the signal and \qqbar\ background.

The likelihood function is
\begin{eqnarray} 
  {\cal L} &=& \exp{\left(-\sum_j Y_j\right)}
  \prod_i^{N}\sum_j Y_j \times \label{eq:likelihood} \\
  &&{\cal P}_j (\mes^i) {\cal P}_j(\xf^i) {\cal P}_j (\DE^i) {\cal P}_j
  (m_{\aone}^i) {\cal P}_j (m_{\Kstar}^i) {\cal P}_j (\calH_{\Kstar}^i, \calH_{\aone}^i),
\nonumber  
\end{eqnarray}
where $N$ is the number of events in the sample, and for each
component $j$ (signal, \qqbar\ background, $b\ra c$ \BB\ background, or charmless
\BB\ background), $Y_j$ is the yield of component $j$ and ${\cal P}_j(x^i)$
is the probability for variable $x$ of event $i$ to belong to component $j$.

\begin{table}[!bp]
\caption{Summary of results for $B^+ \to \aonep \Kstz$. Signal yield $Y$, 
    fit bias $Y_b$, the branching fraction of $\Kstarz \to K^+ \pi^-$
    ${\cal B}(\Kstarz\ra K^+ \pim) $, branching fraction
    $\calB = \calB(B^+ \to \aonep \Kstz) \times \calB(\aonep \to \pi^+ \pi^- \pi^+)$,
    significance $S$ (see text) and upper limit UL. The 
    given uncertainties on fit yields are statistical only, the
    uncertainties on the fit bias include the corresponding systematic uncertainties.}
\begin{center}
\begin{tabular}{ccccccccc} \hline
$Y$                    & $Y_b$             & ${\cal B}(\Kstarz\ra K^+ \pim)$ & $\bfemsix$ & $S$    & UL $(10^{-6})$ \\
\sgline
$\resultY^{+19}_{-17}$ & $\fitbias \pm 14$ & $\frac{2}{3}$    & \rakst     & \sakst & \ulakst  \\
\end{tabular}
\end{center}
\label{tab:results}
\end{table}

We validate the fitting procedure by applying it to ensembles of
simulated experiments with the \qqbar\ component drawn from the PDF,
and with embedded known numbers of signal and \BB\ background
events randomly extracted from the fully simulated MC samples.  By
tuning the number of embedded events until the fit reproduces the yields
found in the data, we find a positive bias $Y_b$, to be subtracted
from the observed signal yield $Y$: the corresponding numbers
are reported in Table~\ref{tab:results}.

In the fitting procedure we allow the longitudinal
polarization $f_L$ to vary, finding the best value 
$f_L = \resultFL$, where the error is statistical only;
systematic uncertainties are not evaluated
and we do not report a measurement on this quantity, 
since the observed signal is not statistically significant. 

\begin{figure}[!H]
\begin{center}
\includegraphics[width=0.95\linewidth]{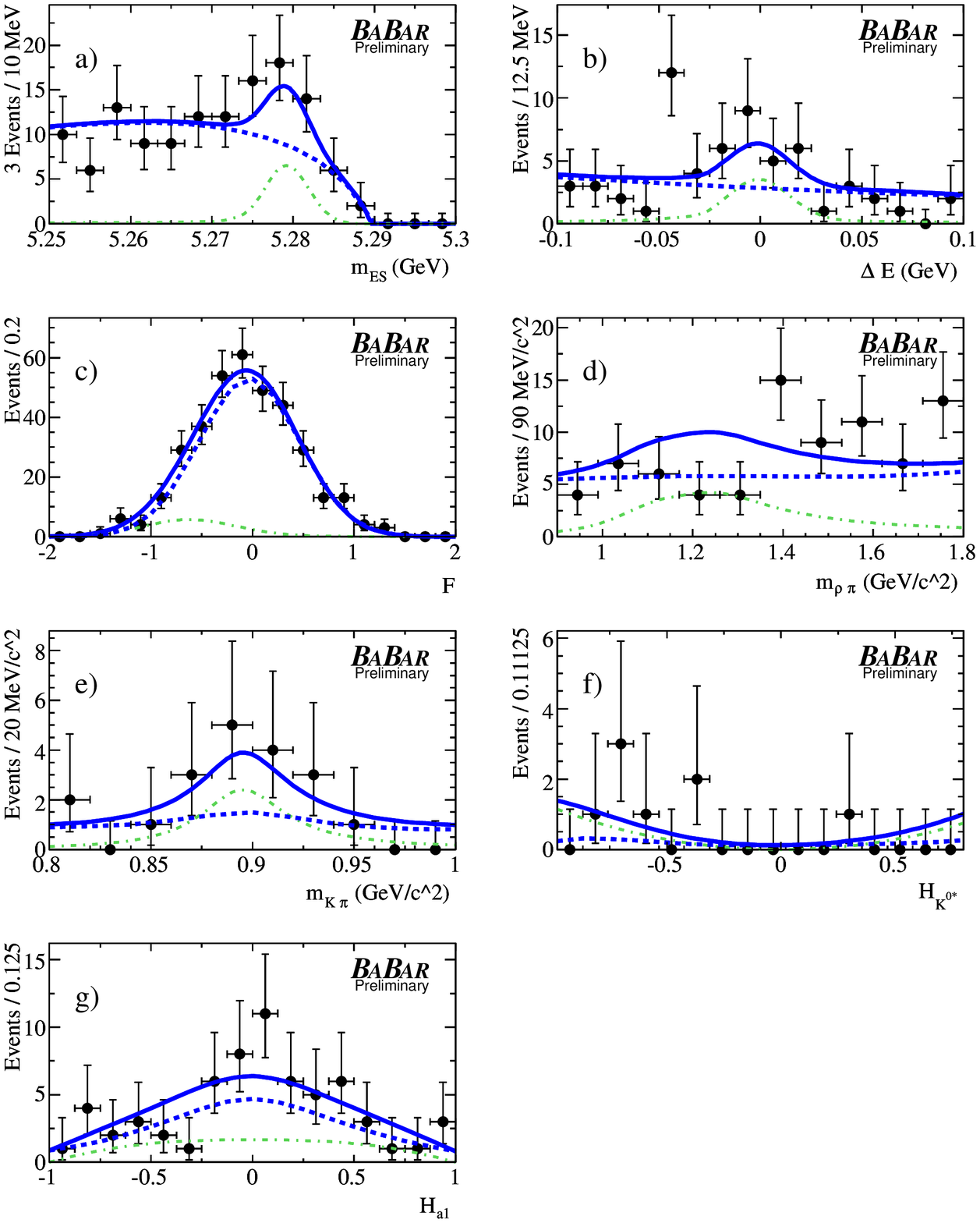}
\caption{Distributions for signal-enhanced subsets (see text)
  of the data projected onto the fit observables for the decay \BcacKstz;
  (a) \mes, (b) \DE, (c) \xf, (d) $m(\rho\pi)$ for the \aone\
  candidate, (e) $m(K\pi)$ for the \Kstar\ candidate, (f) $\calH_{\Kstar}$
  and (g) $\calH_{\aone}$.
  The solid lines represent the results of the fit, and the dot-dashed and
  dashed lines the signal and background contributions, respectively.
  These plots are made with cuts on the ratio of signal to total likelihood.
  With respect to the nominal fit $19\%$ to $46\%$ (depending on the variable)
  of signal events remain.
}
\label{fig:proj_all}
\end{center}
\end{figure}

We compute the branching fraction by subtracting the fit bias from the
measured yield, and dividing the result by the number of produced \BB\
pairs and by the product of the selection efficiency times the branching
ratio for the ${\cal B}(\Kstarz\ra K^+ \pim)$ decay. We assume that the
branching fractions of the \UfourS\ to \BpBm\ and \BzBzb\ are equal,
consistent with measurement \cite{PDG2006}. The efficiency for 
longitudinally and transversely polarized signal events, obtained from MC
signal model, is 12.9\% and 18.6\%, respectively. The results are given
in Table \ref{tab:results}, along with the significance, computed as the
square root  of the difference between the value of $-2\ln{\cal L}$ 
(with additive systematic uncertainties included) for zero signal and
the value at its minimum. In order to obtain the most conservative upper
limit, we assume $f_L=1$ in estimating the branching fraction. In
Figure \ref{fig:proj_all}\ we show the projections of data with PDFs
overlaid. The data plotted are subsamples enriched in signal with
the requirement of a minimum value of the ratio of signal to total
likelihood, computed without the plotted variable.

\section{SYSTEMATIC STUDIES}
\label{sec:Systematics}
Systematic uncertainties on the branching fractions arise from the
imperfect knowledge of the PDFs, \BB\ backgrounds, fit bias, and efficiency.  
PDFs uncertainties not already accounted for by free parameters
in the fit are estimated from varying the signal-PDF parameters within 
their uncertainties. For \Kstar\ resonance
parameters we use the uncertainties from Ref.~\cite{PDG2006} and for the
\aone\ resonance parameters from Ref.~\cite{BaBar_a1pi}.
The uncertainty from fit bias (Table \ref{tab:results}) includes its
statistical uncertainty from the simulated experiments, and half of the
correction itself, added in quadrature.
We vary the \BB\ background component yields by 100\% for charmless
background and by 20\% for the $b\ra c$ \BB\ background.

In the systematic uncertainty we account for a possible
$B^+ \to a_2^+ \Kstarz$ contribution by parameterizing its PDFs on a
dedicated sample of simulated events; for the helicity part of this
component we use the corresponding joint ideal angular distribution
from Ref.~\cite{datta}, as we do for our signal component. We
conservatively assume $B^+ \to a_2^+ \Kstarz$ branching ratio could
be as large as the $B^+ \to \aone^+ \Kstarz$ and vary the
$B^+ \to a_2^+ \Kstarz$ from 0 to 19 events.

The uncertainty from the polarization is obtained by varying $f_L$ within
errors found in studies where $f_L$ was allowed to vary in the fit.   
Uncertainties in our knowledge of the tracking efficiency
include 0.3\% per track in the \B\ candidate.
The uncertainties in the efficiency from the event selection are below 0.6\%.
We determine the systematic uncertainty on the determination of
the integrated luminosity to be 1.1\%.
All systematic uncertainties on the branching fraction are summarized
in Table \ref{tab:sys}.

\begin{table}[!Bth]
\caption{Summary of systematic uncertainties of the determination of the 
         $B^+ \to \aonep \Kstz$ branching fraction.}
\begin{center}
\begin{tabular}{lr}
\hline
Source of systematic uncertainty      &  ~  \\
\sgline
Additive errors (events)              &  ~  \\
~~~$b\ra c$ \BB\ background           &  6  \\
~~~Charmless \BB\ background          &  12 \\
~~~$B^+ \to a_2^+ \Kstarz$ background &  14 \\
~~~\aone\ meson parametrization       &  4  \\
~~~PDF parametrization                &  3  \\   
~~~Variation on $f_L$                 &  2  \\
~~~ML Fit Bias                        &  14 \\
Total additive (events)               &  26 \\
\sgline
Multiplicative errors (\%)                                      & ~ \\ 
~~~Tracking efficiency                                          &  1.2 \\  
~~~Determination of the integrated luminosity                   &  1.1 \\
~~~MC statistics (signal efficiency)                             &  0.6 \\
~~~Differences in the selection efficiency for the \aone\ decay &  3.3 \\  
~~~Particle identification (PID)                                &  1.4 \\  
~~~Event shape restriction ($\costhr$)                          &  1.0 \\  
Total multiplicative (\%)                                       &  4.1 \\  
\sgline
Total systematic error \lbrack\bfemsix\rbrack & $\pm 0.7$  \\
\end{tabular}
\end{center}
\label{tab:sys}
\end{table}

\section{RESULTS}
\label{sec:Results}

We obtain as a preliminary result for the product of branching fractions: 
\begin{eqnarray*}
\calB(B^+ \to \aonep \Kstz) \times \calB(\aonep \to \pi^+ \pi^- \pi^+) =\Rakst \mathrm{,}
\end{eqnarray*}
corresponding to an upper limit of \Ulakst.

Assuming
$\calB(\aone^{\pm}(1260) \to \pi^+ \pi^- \pi^{\pm} )$ is equal to
$\calB(\aone^{\pm}(1260) \to \pi^{\pm} \pi^0 \pi^0)$, and that
$\calB(\aone^{\pm}(1260) \to 3\pi)$ is equal to 100\%,
we obtain:
\begin{eqnarray*}
\calB(B^+ \to \aonep \Kstz) = \RakstprodBR \mathrm{,}
\end{eqnarray*}
corresponding to an upper limit of \UlakstprodBR.
The first error quoted is statistical and the second systematic.
Since the signal significance is \sakst\ standard deviations,
we quote a 90\% confidence level upper limit.

The upper limit from this measurement is, on the one hand, in agreement with the
prediction from na\"{i}ve factorization \cite{Calderon}, and on the other hand, significantly
lower than the QCD factorization estimation \cite{ChengYang}, though not inconsistent with it.

\section{ACKNOWLEDGMENTS}
\label{sec:Acknowledgments}
We are grateful for the 
extraordinary contributions of our \pep2\ colleagues in
achieving the excellent luminosity and machine conditions
that have made this work possible.
The success of this project also relies critically on the 
expertise and dedication of the computing organizations that 
support \babar.
The collaborating institutions wish to thank 
SLAC for its support and the kind hospitality extended to them. 
This work is supported by the
US Department of Energy
and National Science Foundation, the
Natural Sciences and Engineering Research Council (Canada),
the Commissariat \`a l'Energie Atomique and
Institut National de Physique Nucl\'eaire et de Physique des Particules
(France), the
Bundesministerium f\"ur Bildung und Forschung and
Deutsche Forschungsgemeinschaft
(Germany), the
Istituto Nazionale di Fisica Nucleare (Italy),
the Foundation for Fundamental Research on Matter (The Netherlands),
the Research Council of Norway, the
Ministry of Education and Science of the Russian Federation, 
Ministerio de Educaci\'on y Ciencia (Spain), and the
Science and Technology Facilities Council (United Kingdom).
Individuals have received support from 
the Marie-Curie IEF program (European Union) and
the A. P. Sloan Foundation.


\clearpage

\begin{thebibliography}{99}
\bibitem{BaBar_a1pi}
\babar\ Collaboration (B.\ Aubert \etal), \jprl{97}, 051802 (2006); 
\jprl{99}, 261801 (2007).

\bibitem{BaBar_a1K}
\babar\ Collaboration (B.\ Aubert \etal), \jprl{100}, 051803 (2008).

\bibitem{BaBar_b1h}
\babar\ Collaboration (B.\ Aubert \etal), \jprl{99}, 241803 (2007).

\bibitem{conjugate}
Charge-conjugate reactions are implied.

\bibitem{fordFPCP}
W.~T.~Ford, for the \babar\ Collaboration, presented at FPCP, Taipei, Taiwan (2008).

\bibitem{ChengYang}
H.~Cheng and K.~Yang, arXiv:0805.0329v1 (2008).

\bibitem{Calderon}
G.~Calder\'{o}n, J.~H.~Mun\~{o}z, and C.~E.~Vera, \jprd{76}, 094019 (2007).

\bibitem{BABARNIM}
\babar\ Collaboration (B.\ Aubert \etal), \nima{479}, 1 (2002).

\bibitem{geant}
The \babar\ detector Monte Carlo simulation is based on GEANT4
[S. Agostinelli \etal, \nima{506}, 250 (2003)] and EvtGen [D.~J.~Lange,
\nima{462}, 152 (2001)].

\bibitem{PDG2006}
Particle Data Group: Y.-M. Yao \etal, \jpg{33}, 1 (2006) and 2007
partial update for the 2008 edition.

\bibitem{thrust}
A. de R\'{u}jula, J. Ellis, E. G. Floratos and M. K. Gaillard,
\npb{138}, 387 (1978). 

\bibitem{fisher}
R.~A.~Fisher, Annals Eugen. {\bf 7}, 179 (1936).

\bibitem{CB}
M.J. Oreglia, Ph.D Thesis, SLAC-236(1980), Appendix D;
J.E. Gaiser, Ph.D Thesis, SLAC-255(1982), Appendix F and
T. Skwarnicki, Ph.D Thesis, DESY F31-86-02(1986), Appendix E.

\bibitem{PRD04}
\babar\ Collaboration (B. Aubert \etal), \jprd{70}, 032006 (2004).

\bibitem{datta}
A.~Datta \jprd{77}, 114025 (2008).


\end{thebibliography}
\end{document}